\documentclass[preprint,12pt]{elsarticle}
\usepackage{epsfig}
\usepackage{graphics}
\usepackage{graphicx}
\usepackage{amssymb}

\usepackage{lineno}

\journal{Astroparticle Physics}

\begin{document}

\begin{frontmatter}

\title {\bf Multi-wavelength Calibration Procedure for the Pierre Auger
Observatory Fluorescence Detectors}

\author[iafe]{A.C.~Rovero\corref{c_auth}\fnref{conicet}}
\author[csu]{P.~Bauleo}
\author[csu]{J.T.~Brack}
\author[csu]{J.L.~Harton}
\author[csu]{and R.~Knapik}
\author{for the Pierre Auger Collaboration}

\address[iafe]{Instituto de Astronom{\'\i}a y F{\'\i}sica del Espacio
(CONICET), CC 67 Suc 28 (C1428ZAA) Buenos Aires, Argentina}
\address[csu]
{Colorado State University, Department of Physics, Fort Collins CO 80523, USA}

\cortext[c_auth]{Corresponding author. e-mail: rovero@iafe.uba.ar}
\fntext[conicet]{Member of Carrera del Investigador, CONICET}

\begin{abstract}
We present a method to measure the relative spectral response of
the Pierre Auger Observatory Fluorescence Detector. The calibration was done
at wavelengths of 320, 337, 355, 380 and 405~nm using an end-to-end technique
in which the response of all detector components are combined in a single
measurement. A xenon flasher and notch--filters were used as the light source
for the calibration device. The overall uncertainty is 5\%.
\end{abstract}

\end{frontmatter}

\linenumbers

\newpage

\section{Introduction}
\label{sec:Intro}
The Pierre Auger Observatory has been designed to measure Extensive Air
Showers (EAS) initiated by cosmic rays with energies above $10^{18}$~eV.
The Observatory calls for the construction of two large detectors, one in
the southern hemisphere and one in the northern hemisphere, each covering
an area of at least 3000~km$^{2}$ \cite{EA}. The Southern Observatory 
original baseline design in
Malarg\"ue, Argentina, is completed and consists of two detectors,
the Surface Detector (SD) and the Fluorescence Detector (FD). The SD is
composed of 1600 water Cherenkov detectors located on a triangular array
of 1.5 km spacing to measure the EAS secondary particles reaching ground
level. In addition, the UV-nitrogen fluorescence light produced in air is
registered by the FD during dark, clear nights. The FD consists of 24 telescopes
distributed in four buildings, or FD stations, overlooking the SD array.

\begin{figure}[!ht]
\begin{center}
\epsfig{figure=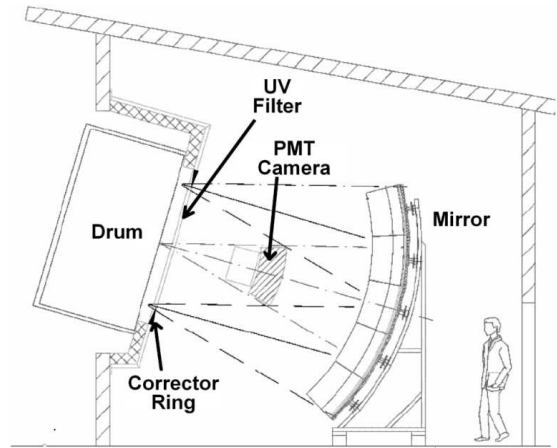,width=8 cm}
\caption{Sketch of the fluorescence detector
showing its main components, together with the calibration light source (drum)
in calibration position. PMT camera supports and drum supports are not shown for clarity.}
\label{fig:FDDrum}
\end{center}
\end{figure}

The energy calibration of data taken at the Pierre Auger Observatory relies
on the calibration of the FD \cite{markus} \cite{spectrum}. A detailed description
of the fluorescence detector can be found elsewhere \cite{FD}. The Auger
FD telescopes  use Schmidt optics. The aperture is defined by a 2.2~m optical
diaphragm. A UV filter covers the aperture and reduces background light
by cutting out all light not in the main part of the nitrogen fluorescence
spectrum ($\sim$300 - 400 nm). It also provides ambient isolation, which
allows for temperature controlled operation of the telescope and prevents
dust from entering the optical system. Spherical aberrations are reduced by a
Schmidt corrector annulus covering only the outer portion of the aperture. Light
is concentrated by a 3.5~m $\times$ 3.5~m tessellated spherical
mirror into an array of 440 hexagonal photomultipliers (PMTs), referred
to as ``pixels'', with a field of view of 1.5 $\deg$ each. At the focal plane,
light concentrators approximating hexagonal Winston cones reduce dead spaces
between PMTs. The pixel array is referred to as a ``camera''. In
Fig.~\ref{fig:FDDrum} we show the main components of the FD telescope.

To calibrate the FD three different procedures are performed \cite{rob}: the
relative, the absolute and the multi-wavelength calibrations. Relative
calibration is performed at least at the beginning and at the end of every
observing night. It is based on uncalibrated but stable light sources that
illuminate the camera from three positions upstream in the optical system,
tracking the nightly response variations of the whole system \cite{calA}. The
absolute calibration is made by an end-to-end technique, using a calibrated portable
light source in front of the telescope aperture, which calibrates the combined
effect of each component in a single measurement at a single wavelength, 375 nm.
The light source has been designed to uniformly illuminate all 440 pixels in
a single camera simultaneously and is referred to as the ``drum'' because of
its appearance. It is a cylinder of 2.5 m diameter and 1.4 m deep, with one
Teflon$^{\circledR }$ face, and internally laminated with Tyvek$^{\circledR }$ 
(see Fig.~\ref{fig:drum}). When used for absolute calibration of FD telescopes,
a UV LED is placed inside a small Teflon diffuser inside the drum, and surrounded
by other diffusive pieces in such a way that the face is uniformly illuminated. 
The procedure to calibrate the drum at the laboratory has been
outlined elsewhere for the prototype \cite{jeff} and for the current version of
the drum \cite{pune}. Absolute calibration of FD telescopes is performed
typically twice a year to follow long term variations of the system response. 
Finally, a ``multi-wavelength calibration'' procedure determines the spectral
response of the system as a function of photon wavelength. This is a relative
measurement, normalised to the absolute calibration at 375 nm. The multi--wavelength
calibration is needed not only for correct event reconstruction but also to
correlate with the results of alternative absolute calibrations performed at
different wavelengths using lasers. Changes in the spectral response of the
FD are not expected to occur in the short term, thus the frequency for
evaluating this dependence is planned to be less than once per year.

In this work, we describe the procedure for multi-wavelength calibration of
fluorescence telescopes using an end-to-end technique similar to that used for the
absolute calibration. We describe the initial spectral dependence function used by the
Auger Observatory in section \ref{sec:Piecewise}. The new light source used in the
procedure is described in section \ref{sec:Hardware} and its characterisation
in section \ref{sec:measurements}. The FD-telescope wavelength response
and a discussion of uncertainties are presented in section \ref{sec:response}.

\section{Piecewise spectral response of the fluorescence detector}
\label{sec:Piecewise}
The spectral response of the FD originally used by the Auger Observatory was
assembled from the efficiencies of the individual telescope components. The
individual efficiencies were obtained from statements by the component manufacturers
or, in some cases, as measured by members of the Pierre Auger Collaboration
\cite{piecewise}. The elements considered for the spectral
response were the UV filter and corrector ring transmission, the mirror
reflectivity, and the PMT quantum efficiency. The overall wavelength response is
dominated by filter and PMT effects. We call this piece-wise curve $PW(\lambda)$
and show it in section \ref{sec:response} to compare it with results in
Fig.~\ref{fig:results}.

Regardless of the accuracy of the measurements on each telescope component,
the fact that the response of the system was not measured as a whole indeed
introduces uncertainties. We can calculate, for example, that the corrector
ring transmission does not affect all the incoming photons but only approximately
half of them, when camera shadowing effects are included. Also, the reflectivity
of the light concentrators in the camera are not taken into account at all,
even though a third of the light getting to the pixel is reflected by them.
There are other less important considerations such as the fact that mirrors for
the FD telescopes are made using two different techniques by two different
manufacturers \cite{mirrors}. One of them is a machined aluminium alloy with a
protecting layer of Al$_2$O$_3$ at its reflecting surface. In the second
technique polished glass is aluminized and the reflecting surface is covered
by a layer of SiO$_2$. Both types of mirrors have similar reflectivities as
measured at 370 nm \cite{mirrors}. However, considering that they use different
dielectric coatings on different materials, the reflectivity at other
wavelengths may vary.

To assure that the Pierre Auger Observatory is using the right spectral response
of its FD telescopes, the decision was made to adapt the end-to-end procedure
used for absolute calibration to directly measure this function.

\section{Multi-wavelength light source}
\label{sec:Hardware}
\begin{figure}[ht]
\begin{center}
\epsfig{figure=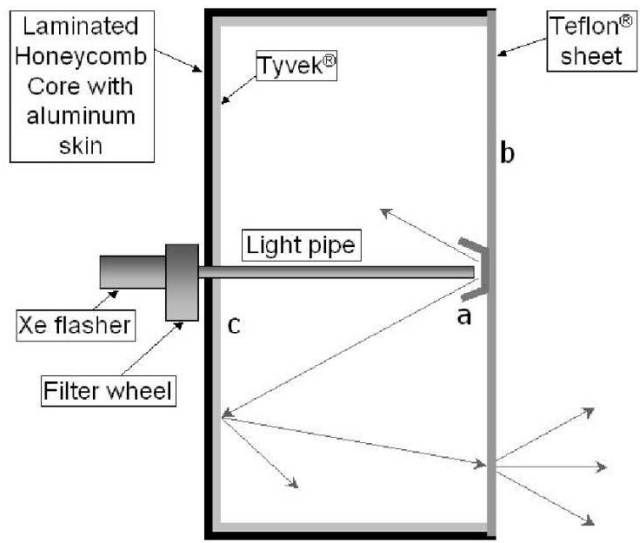,width=10 cm}
\caption{Drum section showing the main components of
the light source. The diffusively reflecting surfaces a) and b) are Teflon, while c) is Tyvek}
\label{fig:drum}
\end{center}
\end{figure}

To enable multiple wavelength measurements, the LED used for absolute
calibration was removed and a light pipe was installed between the
Teflon diffuser and the back of the drum, where new light sources could be
mounted (see Fig.~\ref{fig:drum}). A xenon flasher is mounted at the end of
the pipe at the back of the drum. The xenon flasher\footnote{RSL-2100 series
xenon flasher - Perkin Elmer, www.perkinelmer.com} provides 0.4 mJ optical
output per pulse covering a broad UV spectrum, in a time period of a few
hundred nanoseconds. To select a desired wavelength,
notch-filters\footnote{Standard UV bandpass filters, Andover Corporation, 
www.andcorp.com} are mounted in a filter wheel attached to the end of the pipe.
A focusing lens at the filter wheel output maximises the intensity through
the filter wheel and into the light pipe.
Notch-filters were chosen at five wavelengths inside the range of the FD UV
filter located at the telescope aperture. According to the manufacturer, the
filter transmissions are centred at 320, 337, 355, 380 and 405~nm, with a
FWHM $\approx$~15~nm.

\section{Characterization Measurements}
\label{sec:measurements}
The drum relative intensity is measured at each wavelength in a dedicated
calibration laboratory at the Observatory. An auxiliar PMT, the ``lab-PMT'',
is used to measure the light intensity of the drum for a given notch-filter.
Quantum efficiency (QE) of the lab-PMT and FD spectral response have significant
variations within the range of wavelengths where the notch-filters transmit.
To understand the corrections to be applied due to these variations we have
performed measurements described in the following sections.

\subsection{Notch-filter transmission scan}
\label{sec:filters}
\begin{figure}[ht]
\begin{center}
\epsfig{figure=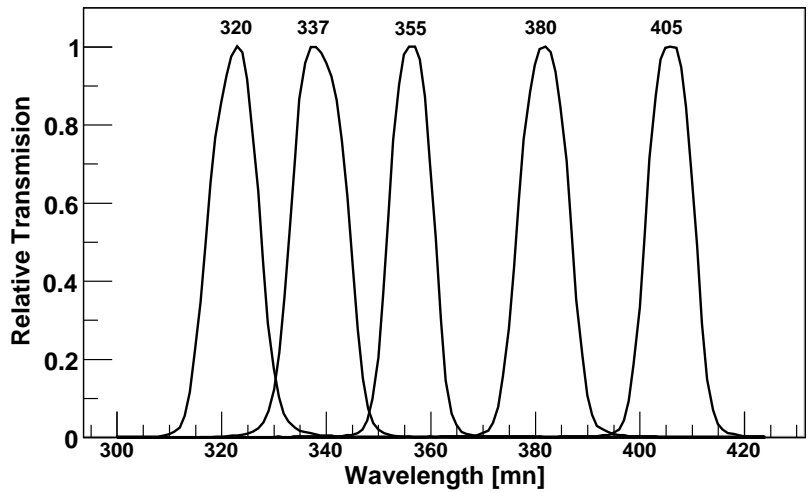,width=12 cm}
\caption{Relative transmission of the notch-filters used
for multi-wavelength calibration. The nominal wavelength is indicated for each filter}
\label{fig:filters}
\end{center}
\end{figure}

Notch-filters listed in section \ref{sec:Hardware} were selected at 5
wavelengths either near the main emission bands of nitrogen \cite{n2} used
for EAS fluorescence detection (320, 355 and 405~nm), or to match the laser
wavelengths (337 and 355~nm) used in previous absolute calibration cross checks
using roving lasers \cite{rob}, or near the 375 nm LED single--wavelength absolute
calibration. Precise measurements of the transmission characteristics of each
notch-filter are needed.  The filters were scanned using a monochromator with a
broadband deuterium light source. A photo-diode of known wavelength dependence
\cite{nist} was used at the monochromator output to detect the transmitted light
as a function of wavelength in 2 nm steps. The results are shown in
Fig.~\ref{fig:filters} where the spectral shape of the deuterium light source
and the response of the photo-diode have been deconvolved. We assign labels
of $f_i(\lambda)$ to the curves in the figure and $\lambda_i$ to the central
wavelength of each of them, where $i = 1,5$ indicates one of the five
notch-filters (from 320 nm to 405 nm).
A scan from the manufacturer was available for one of the filters (337~nm),
and it was found to be in good agreement with our scan. Some asymmetries
were found in the transmission curves (320 and 337 nm) that make significant
differences when applying filter corrections.

\subsection{Quantum efficiency of the lab-PMT}
\label{sec:pmtqe}
We measured the relative quantum efficiency of the lab-PMT in 2~nm steps using
the deuterium light source and the monochromator. We directed the monochromator output
through a thin Teflon diffusor and into a dark box containing the lab-PMT and a
NIST-calibrated photo-diode\footnote{UV100 photodiode, UDT Sensors, Inc, Hawhorne, CA USA}
\cite{nist}. Using the small photo-diode we verified that the beam was uniform
laterally at a level of 0.5\% over an area larger than the PMT photochathode. The first step in
the QE measurement was to scan the monochromator and measure the relative output
intensity in photons at each wavelength, using the photo-diode and it's known
calibration.  Then, using an iris to limit the intensity and prevent PMT saturation,
we rescanned the source and measured the PMT current. The PMT relative QE as a function of
wavelength, $QE(\lambda)$, is the ratio of these two scan results at each wavelength.

\begin{figure}[ht]
\begin{center}
\epsfig{figure=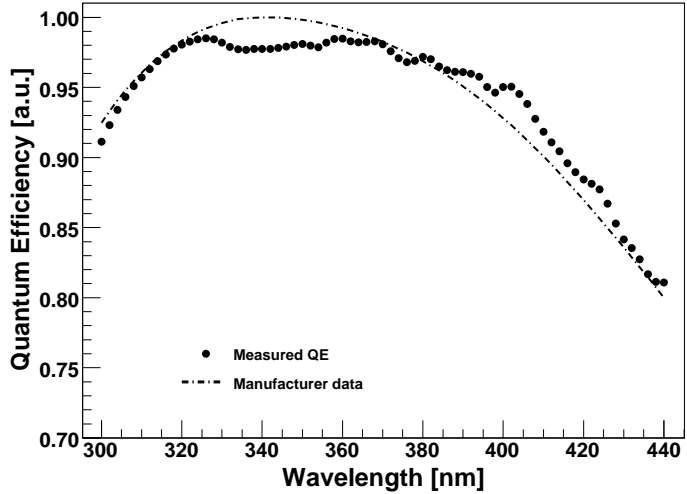,width=10 cm}
\caption{The measured quantum efficiency of the lab-PMT
used in this work, and the typical QE from the manufacturer's specification sheet}
\label{fig:qe}
\end{center}
\end{figure}

The measured QE, shown in Fig.~\ref{fig:qe}, is in agreement with the average
photocathode QE provided by the manufacturer\footnote{78 mm, 9265B PMT, made by Electron
Tubes, www.electrontubes.com}, within the measurement uncertatities of 2.5\%, which are 
discussed in section~\ref{sec:uncertainties}.

\subsection{Drum Intensity at five wavelengths}
\label{sec:intensity}
With the notch-filter wheel mounted on the drum, the relative intensity
of the drum surface for each wheel position depends on the xenon source
intensity at the transmitted wavelengths, the notch-filter transmission
and losses in the light pipe. No wavelength shifting of photons of Teflon
and Tyvek materials has been observed in our independent laboratory
measurements. Ideally, to measure the drum intensity for each wavelength,
one would use the drum surface as input to the monochromator and scan the
full spectrum for each notch-filter. In practice this is precluded by
the low drum intensity. Instead, we make a single measurement of the integrated
drum intensity for each notch-filter using the lab-PMT. For each notch-filter
we find a value for $C_i$, the centroid of the histogram of the PMT response
to 1000 xenon flasher pulses.
These centroids are proportional to the drum intensity for each
notch-filter once corrections have been made for variations in lab-PMT
QE. Ignoring common constants, $\int \Phi_i(\lambda)~QE(\lambda)~d\lambda = C_i$,
where $\Phi_i(\lambda)$ is the brightness of the drum surface for the
notch-filter $i$ as a function of wavelength. Then, since the distribution
of drum photons is the convolution of the known xenon flasher spectrum,
$Xe(\lambda)$, and the corresponding notch-filter, we use
$\Phi_i(\lambda) = k_i f_i(\lambda)~Xe(\lambda)$ and adjust $k_i$ to match
the integral above. With this last process all five values of drum brightness
and their wavelength distributions are known.

The quantity $\int \Phi_i(\lambda)~d\lambda = \Phi_i$ is proportional to the
real total photon flux being emitted by the drum surface. All the wavelength
independent properties (PMT gain, electronic conversion, etc.) are left out
because they will cancel in the end when the relative values are computed.
We also note that $\Phi_i$ is not significantly different from the value
it would have if the function $QE(\lambda)$ was totally flat within the
notch-filter range. Then, in practice,

\begin{equation}
\int \Phi_i(\lambda)~d\lambda = \Phi_i \simeq C_i/QE(\lambda_i)
\label{eq0}
\end{equation}

\section{Fluorescence Detector response to drum}
\label{sec:response}
\begin{figure}[ht]
\begin{center}
\epsfig{figure=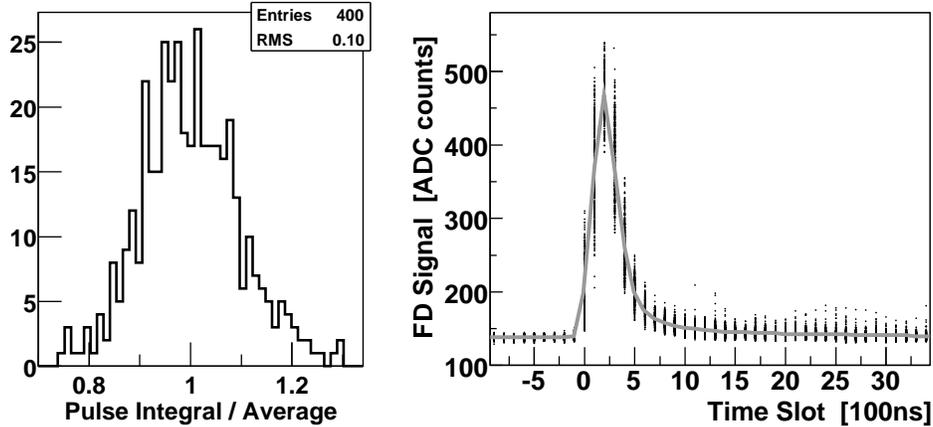,width=13cm}
\caption{FD single pixel response to drum pulses.
The signal of pixel number 220 of telescope 4 with drum at 355 nm is shown. Left:
Distribution of pulse integrals related to the average integral. Right: Pulses as
registered by the FD. Dots are the values for 400 individual pulses. The grey solid
line is the average}
\label{fig:xenonpulse}
\end{center}
\end{figure}

For testing the procedure we use results of measurements made at one FD
telescope. In August 2006 we mounted the drum with xenon flasher and filter wheel
at the aperture of telescope 4 at the Los Leones FD--building. A series of 400 xenon
flashes illuminated the camera, and we found the
average integrated pulse for each pixel. In Fig.~\ref{fig:xenonpulse} we
show the response of one FD pixel and the distribution of pulse integrals
for the same pixel. The pulse shape from the drum with the xenon source is
irregular and varies from pulse to pulse, but the total output energy is
consistent as indicated by the $<$10~\% RMS of the integral distributions.
We obtained the average charge from the distributions of those 400 pulse
integrals, $I_{i,j}$, for each notch-filter $i$ and 
each pixel $j$. Typical statistical uncertainty for these values is $<$0.5~\%.

\begin{figure}[ht]
\begin{center}
\epsfig{figure=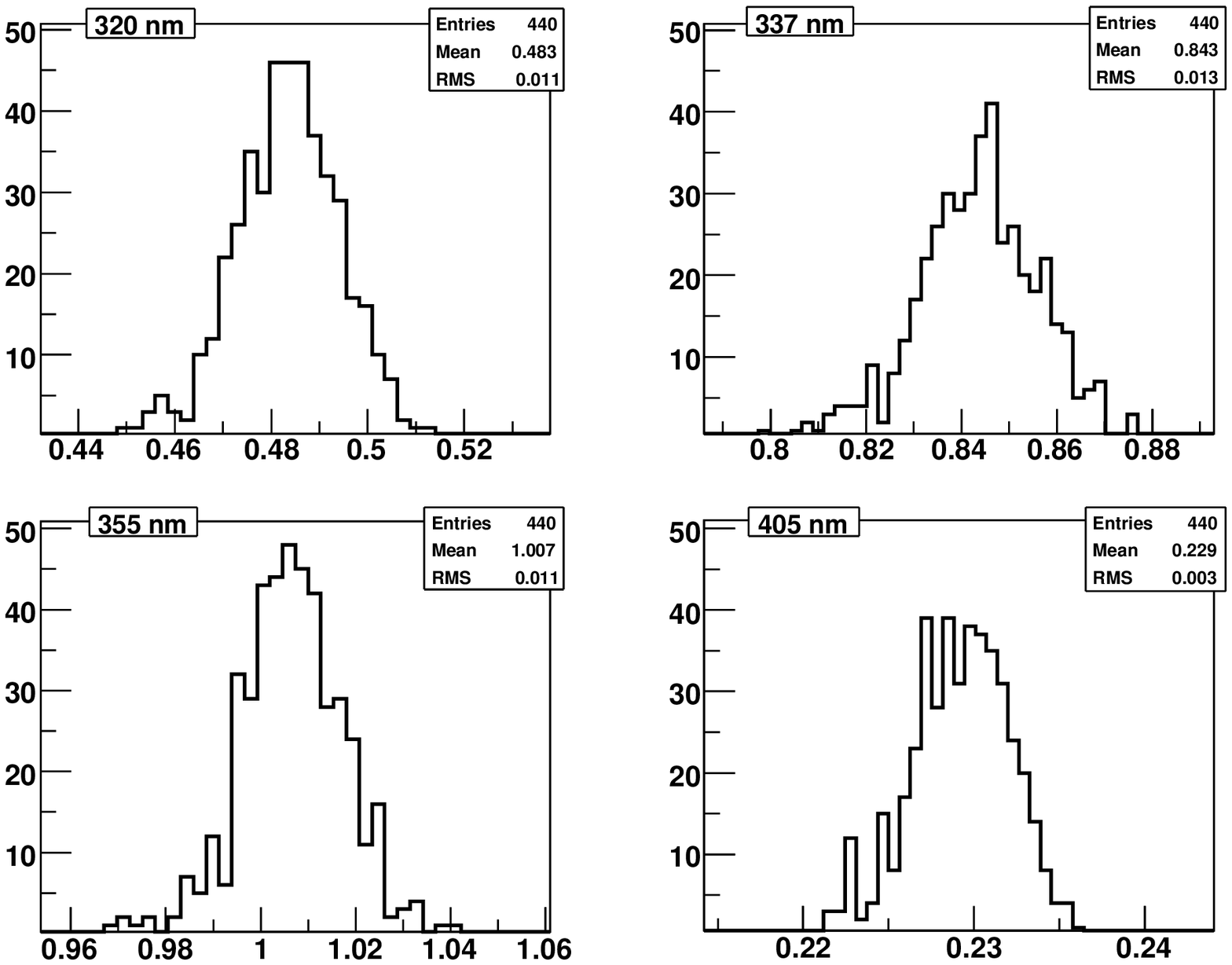, width=14 cm}
\caption{Distribution of relative FD response
for each notch-filter. The response relative to 380 mn of all pixels in one telescope is shown}
\label{fig:PMTresponse}
\end{center}
\end{figure}

The relative wavelength-dependent FD response for pixel $j$ is then the ratio
of the integrated ADC response of the FD to the relative number of photons at
the aperture. We call this value $R_{i,j}$ so that $R_{i,j} = I_{i,j} / \Phi_i$.
A considerable dispersion of $R_{i,j}$ values is expected as pixels have
different gains and amplifications. However, we are only interested in relative
wavelength responses so, all values are normalized to the response to the
380 nm notch-filter, or $i = 4$. Thus $R_{i,j}^{rel} = R_{i,j} / R_{4,j}$.
In Fig.~\ref{fig:PMTresponse} the distribution of these relative responses
for pixels in the measured telescope is shown for each notch-filter, except
the one at 380 nm as this would be the identity.
The histograms shown for each notch-filter in Fig.~\ref{fig:PMTresponse} have 
relatively low dispersions, ranging from 1.1\% to 2.2\% RMS. We consider that
the relative response for each PMT in the camera is well represented by the
average of those distributions, so only one value for each notch-filter is
taken. We call these values $R_i^{rel}$, where $i$ identifies the filter.

In Table~\ref{tab:rel} we show the $R_i^{rel}$ values obtained for the telescope
as evaluated from the distributions on Fig.~\ref{fig:PMTresponse}.
Uncertainties to these values are discussed in next sections.
The resulting wavelength dependent efficiency is shown in Fig.~\ref{fig:results}
where the fitting was done adjusting the piecewise function to the five measured
points $R_i^{rel}$. The piecewise response in its original form is also shown in the
figure for comparison purposes. The fit has been corrected for notch-filter width
effects, as described in next section. 

\begin{table}[ht]
\begin{center}
\begin{tabular}{|c|cc|}
\hline
        Wavelength        & \multicolumn{2}{c|}{ $R_i^{rel}$}   \\ 
\hline
	    (nm)          &      Mean     &  Corrected    \\
\hline
            320           &      0.483    &  0.409      \\
	    337           &      0.843    &  0.832      \\
	    355           &      1.007    &  0.998      \\
	    380           &      1.000    &  1.044      \\
	    405           &      0.229    &  0.243      \\
\hline      
\end{tabular}
\caption{Average relative FD response measured for each
notch-filter. Values are taken from the distributions of Fig.~\ref{fig:PMTresponse}.
The column on the right have the values after the correction for notch-filter
width effects as described in section \ref{sec:filter}}
\label{tab:rel}
\end{center}
\end{table}

\subsection{Notch Filter Width Effects and fitting procedure}
\label{sec:filter}
In section \ref{sec:intensity} we described how the drum centroids for
five wavelengths were measured and corrected for the lab-PMT quantum
efficiency to get the relative drum intensities. In that process the change
in the QE was taken into account by considering the distribution of photons
for each notch-filter. Then, in section \ref{sec:response}, we described how
the FD responses to those intensities were measured for one FD telescope. In
this last process the notch-filter width effect was not taken into account.
Because the overall FD response is not flat in the $\sim$15~nm FWHM
range of each filter, this uncorrected result is biased toward the region of the
filter corresponding to higher FD response.

To correct for this effect we follow a similar procedure as in section
\ref{sec:intensity}. The process at the telescope is:

\begin{equation}
\frac{\int \Phi_i(\lambda)~FD(\lambda)~d\lambda}{\int \Phi_i(\lambda)~d\lambda}
= R_i^{rel}
\label{eq1}
\end{equation}

where $FD(\lambda)$ is the FD relative response as a function of wavelength.
Again, we are not including wavelength independent constants as this is a
relative measurement procedure. Note that if the notch-filter transmission,
$f_i(\lambda)$, were delta functions at $\lambda_i$, we would have
$FD(\lambda_i) = R_i^{rel}$. To solve the integral of equation \ref{eq1} and
find the unknown function $FD(\lambda)$ we made the assumption that the FD
response is a piecewise-like function, as described in section \ref{sec:Piecewise}.
This choice is required because the five measurements in the wavelength range of
the FD response are not sufficient to fully characterise a generic function.
Then, it is reasonable to adjust the original piecewise function, $PW(\lambda)$,
to what was measured. This assumption also implies
that we have taken the FD response as null beyond the wavelength of the FD
system. Particularly, we take $FD(280~nm) = FD(425~nm) = 0$, which have
been verified by measuring the UV-filter transmission in our laboratory.

The procedure of adjusting the fitting function to the measured points was as
follows. For each notch-filter we take $FD(\lambda) = h_i Fit(\lambda)$,
where $Fit(\lambda)$ is the fitting function and $h_i$ a parameter for filter
$i$. As a first guess the piecewise function is taken so,
$Fit(\lambda) = PW(\lambda)$. Equation \ref{eq1} is evaluated and the parameter
$h_i$ found to match the value $R_i^{rel}$. Once all five $h_i$ are
found, a new fit is done by interpolating the points $h_i Fit(\lambda_i)$ with
a piecewise-like function, where now $i$ runs for seven points, including the
null extremes.  Between these points, the curve is adjusted by a linear
interpolation of the adjustments at the surrounding points. Finally, this
last fit is taken as a new $Fit(\lambda)$ function and the process starts again
until all five values $h_i$ are the identity.

\begin{figure}[ht]
\begin{center}
\epsfig{figure=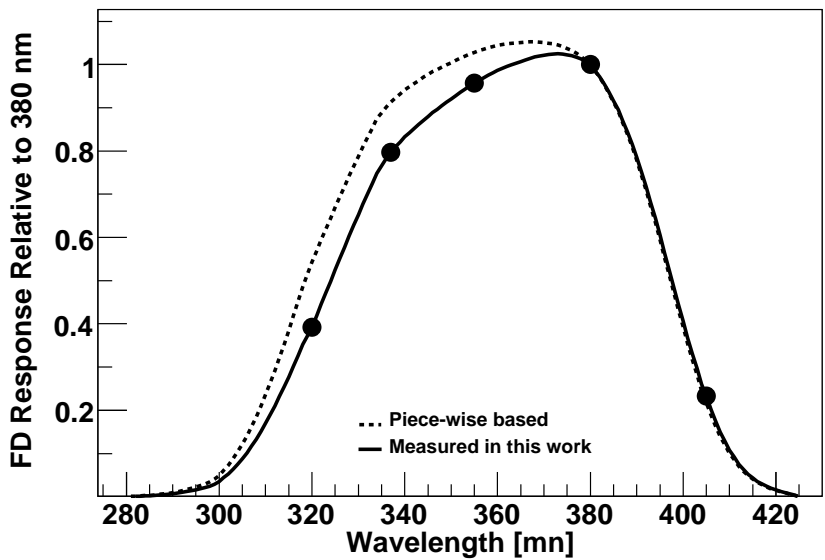,width=12 cm}
\caption{The measured Fluorescence Detector wavelength
response compared with a curve generated in a piecewise fashion from manufacturer's data
for each FD component. The corrected measurements at the 5 wavelengths are shown; the
solid curve is constrained to pass through those points}
\label{fig:results}
\end{center}
\end{figure}

The final result of the iteration procedure, $Fit(\lambda)$, is the $FD(\lambda)$
that fulfils the integral in equation \ref{eq1} for all five measured points.
This curve is shown in Fig.~\ref{fig:results}, normalised to the value at 380~nm.
In the same figure we also show the original piecewise function. A decrease in
the spectral response compared to the piecewise response is observed at shorter
wavelengths, the largest difference of $\approx$~-28\% comes at 320~nm. The corrected
$R_i^{rel}$ values are shown in Table~\ref{tab:rel}.

\subsection{Uncertainties}
\label{sec:uncertainties}

Contributions to the uncertainty include those from measurements in
the laboratory of the drum intensity, and from measurements at the
telescope of the FD response.  The overall result reported here is a
relative measurement, and consequently many factors cancel
particularly systematics related to laboratory setup.\\

The determination of drum intensity at each wavelength includes
measurements in the laboratory of the centroid of the lab-PMT response
to pulsed drum illumination, and the relative QE of the lab-PMT.
The uncertainty in the lab-PMT drum response centroid, $C_i$
in equation \ref{eq0}, is estimated to be 1 channel in the ADC converter plus
the statistical uncertainty on the mean of the 1000 xenon pulse
distribution. The 1 channel uncertainty is a systematic effect,
related to repeatability, and has more relevance for wavelengths where
the drum brightness is low. The second column in Table~\ref{tab:error}
indicates the uncertainties in $C_i$.

\begin{table}[ht]
\begin{center}
\begin{tabular}{|c|c|c|ccc|}
\hline Wavelength &  $C_i$  & $\Phi_i$  &              & $R^{rel}_ i$ (\%) &        \\
          (nm)    &  (\%)   &  (\%)     & Statistical  &  Systematic       & Total  \\
\hline
       320  &  0.6  &  2.6  &  2.2  &  4.1  &  4.7  \\
       337  &  1.0  &  2.7  &  1.5  &  4.2  &  4.5  \\
       355  &  2.0  &  3.2  &  1.1  &  4.5  &  4.6  \\
       380  &  2.0  &  3.2  &  n/a  &  0.0  &  0.0  \\
       405  &  0.3  &  2.5  &  1.3  &  4.1  &  4.3  \\
\hline

\end{tabular}
\caption{Sources of uncertainties and their values for this work (see
section~\ref{sec:uncertainties}).}
\label{tab:error}
\end{center}
\end{table}

Uncertainties related to measurement of the relative QE of the lab-PMT
include those from lab-PMT response, photodiode current during monochromator
scans, the calibration of the photodiode at each
wavelength, and systematics in the laboratory setup.  The uncertainty
in the monochromator wavelength has been measured with a N2 laser
light source to be less than $\sim$0.25 nm, and no contribution to the
overall uncertainty has been included for this effect.

For the PMT QE scans, currents from the photodiode and the lab-PMT
were measured with electronics based on an integration chip with
linearity of 0.005\% \cite{ivc102}.  Effects of connectors and cabling
between the detectors and the integration chips are expected to
dominate any nonlinearities.  While these effects are expected
to be small, we assign an overall uncertainty of 2\% to current
measurements, based on our experience measuring absolute currents
using similar configurations.  For this relative measurement, the
experimental configuration remained unchanged during measurements at
different wavelengths, allowing some readout systematics to cancel in
the final ratios. The uncertainties in the absolute response of the
photodiode, as calibrated by NIST, are documented as k=2 expanded
uncertainties \cite{nist}.  Assuming normal distributions for the
contributing factors, these can be interpreted as 1-sigma
uncertainties varying in the range from 0.55 to 0.95\% for wavelengths
between 300 and 400 nm, and 0.4\% or less for those between 405 and
450 nm.  Monochromator output beam uniformity has been measured in the
PMT-photodiode region to be better than 0.5\% across a 10 cm region
perpendicular to the beam.  No contribution to the overall uncertainty
has been included for beam non-uniformity.  The stability of the
system, including the lab-PMT response, was tested by performing
measurements multiple times to find that repeatability is well within
1\%.  As a result of these effects we consider an overall uncertainty
of 2.5\% for the relative QE of the lab-PMT.  Combination of the
uncertainties of $C_i$ and QE give the uncertainty in the drum
brightness, $\Phi_i$, as listed in Table~\ref{tab:error}.\\

The camera--averaged response of pixels to each of the five
wavelengths are described in section~\ref{sec:response}.  To evaluate
the uncertainties in these responses we consider the
systematics coming from drum fluxes, $\Phi_i$ as discussed above, and
the widths of the distributions of responses for the 440 pixels in the
camera, as taken from the distributions of
Figure~\ref{fig:PMTresponse} and shown in the fourth column of
Table~\ref{tab:error}. For a given pixel $j$,
$R^{rel}_{ij}=R_{ij}/R_{4j}$, four quantities have to be considered:
two integrals, $I_{ij}$ and $I_{4j}$, and two drum fluxes, $\Phi_i$
and $\Phi_4$, as defined in previous sections.  Uncertainties in the
integrals are purely statistical and have been evaluated to be
$<$0.5~\%.  The combination of these four values gives the total
uncertainty shown in the fifth column of Table~\ref{tab:error}.
Values at different wavelengths are independent and are added in
quadrature, except at 380 nm where systematic uncertainties are null
by definition. The last column in Table~\ref{tab:error} shows the
combined uncertainty for the two sources.\\

Finally, we consider uncertainties resulting from the fitting
procedure incorporating the effects of the notch-filter widths, as
described in section \ref{sec:filter}.  The shape of the piecewise
calibration curve, used as the initial fitting function, is dominated
near 300 nm by the fall-off of FD PMT quantum efficiency at shorter
wavelengths, and by decreasing transmission of the UV filter through
the FD aperture above 400 nm.  We have tried several shapes for this
initial function and found that any reasonable choice incorporating
these two features leads to near-identical results. No related
uncertainty is included.

When applying the notch-filter width correction we use the relative
transmission of the 5 notch filters, measured as described in
section~\ref{sec:filters}.  Associated uncertainties are 0.5\% at each
wavelength in the scans.  Good correlation with the manufacturer's
scan available for one filter supports this value of the uncertainty.
For an independent evaluation of the systematics related to filter
shape, we took the wavelength with the largest filter-width effect,
320 nm, and changed the filter shape to an extreme-case scenario of a
step function.  We included a maximum spread in intensity between two
wavelengths by assuming extremes of the photodiode uncertainties at
the respective wavelengths. We then repeated the notch filter
correction calculation.  The corrected value of $R^{rel}_{1}$ given in
Table~\ref{tab:rel} changed by -0.5~\%, supporting the assesment
above.

Based on the discussions above, we assign an overall total uncertainty
of 5~\% to the measurements reported here.  The main contributions
come from the uncertainties on the relative QE of the lab-PMT and the
measured relative drum fluxes at each wavelength.

\section{Conclusions}
\label{sec:conclusions}
The method for measuring the relative wavelength-dependence response of the
Pierre Auger fluorescence detector has been tested in one telescope. Within
uncertainties we can say that a multi-wavelength calibration for each PMT in a
given camera is not necessary as the dispersion around the average is of the
order of few percent. 

This result indicates lower FD efficiency at shorter wavelengths when compared
to a curve constructed in a piecewise manner from manufacturer efficiency
specifications for the individual elements of the system. The piecewise curve
was adjusted to the measured values and the result is currently used in the
Auger event reconstruction software.

\newpage


\begin{thebibliography}{00}

\bibitem{EA} J. Abraham [Pierre Auger Collaboration], NIM A 523 (2004) 50-95.

\bibitem{markus} M. Roth [Pierre Auger Collaboration], Proceeding 30$^{th}$
ICRC, Merida (2007).

\bibitem{spectrum} J. Abraham [Pierre Auger Collaboration], Physical Review Letters
101 (2008), 061101.

\bibitem{FD} V. Verzi [Pierre Auger Collaboration], Nucl. Phys. B (Proc. Suppl.)
165 (2007) 37.

\bibitem{rob} R. Knapik [Pierre Auger Collaboration], Proceeding 30$^{th}$ ICRC,
Merida (2007).

\bibitem{calA} C. Aramo [Pierre Auger Collaboration], Proceeding 29$^{th}$ ICRC,
Pune (2005), 8, 101.

\bibitem{jeff}  J.~T.~Brack, {\em et al.}, Astropart. Phys. 20 (2004) 653.

\bibitem{pune} P. Bauleo [Pierre Auger Collaboration], Proceeding 29$^{th}$
ICRC, Pune (2005), 8, 5-8.

\bibitem{piecewise} J.~A.~J.~Matthews [Pierre Auger Collaboration], Proceeding
SPIE Astronomical
Telescopes and Instrumentation, Waikoloa , Hawaii (2003), 4858, 121-130.

\bibitem{mirrors} G. Matthiae [Pierre Auger Collaboration], Proceeding 27$^{th}$
ICRC, Hamburg (2001).

\bibitem{n2} M. Nagano, {\em et al.}, Astropart.Phys. 20 (2003) 293.

\bibitem{nist} T.C. Larson, S.S. Bruce, and A.C. Parr, ``Spectroradiometric
Detector Measurements'',
National Institute of Standards and Technology, Calibration Program,
Gaithersburg, MD 20899-2330,
Special Publication 250-41, 1998.

\bibitem{ivc102} Burr-Brown Corporation, chip ivc102; http://www.burr-brown.com/


\end{thebibliography}
\end{document}